\pdfoutput=1 

\documentclass[final,1p,times]{elsarticle} 

\usepackage{graphicx} 
\graphicspath{ 
  {/home/lindenle/figs/},
  {./figures/}}

\newcommand{\jpsi}{{{\rm J}/\psi}}

\usepackage{color}

\usepackage{amssymb} 
\usepackage{amsthm} 

\usepackage{lineno} 
\usepackage{wrapfig}

\journal{Nuclear Physics A} 
\begin{document} 

\begin{frontmatter} 


\title{From production to suppression,\\%
  a critical review of charmonium measurements at RHIC.}

\author{L. A. Linden Levy}

\begin{abstract}
 
Charmonium suppression in hot and dense nuclear matter has been argued
to be a signature for the production of the quark gluon plasma
(QGP). In order to search for this effect in heavy ion collisions one
must have a clear understanding of all the factors that can contribute
to such a suppression. These may include shadowing of the partons in a
nuclear environment, breakup of a correlated $c-\bar{c}$ pair as it
traverses the nuclear fragment, suppression of feed-down from higher
mass states as well as other initial state interactions. In order to
disentangle these effects one must measure charmonium production rates
in both proton+proton (p+p) and proton+nucleus (p+A) collisions. The
p+p collisions serve as a baseline for searching for suppression
compared to binary scaling predictions, allow one to quantify the
amount of feed-down from higher states as well as serve as a tool to
distinguish between different theoretical calculations for charmonium
production mechanisms. In order to quantify nuclear effects it is also
necessary to study charmonium production in p+A collisions where the
temperature and density of the system are low compared to a heavy ion
collision. These measurements allow one to determine the influence of
nuclear shadowing and breakup in ``cold'' nuclear matter which can be
extrapolated to heavy ion collisions in order to determine the amount
anomalous suppression. Of course, extrapolations that rely on a model
based technique depend heavily on the assumption of a production
mechanism, a fact that reinforces the importance of the p+p
measurements. The PHENIX and STAR experiments at Brookhaven National
Laboratory have measured charmonium production in p+p, d+Au, Au+Au and
Cu+Cu collisions at $\sqrt{s_{NN}}$ = 200 GeV for both forward and mid
rapidities.  I will present a review of the latest measurements from
both experiments with an emphasis on what we have and can still
learned from them about charmonium production and suppression with
these experimental apparatuses.

\end{abstract}

\address{University of Colorado, 390 UCB, Boulder Co, 80309, USA}

\end{frontmatter} 


\section{Why the $\jpsi$}
\label{sec:introduction}
The heavy nature of charmonium ($c\bar{c}$) allows one to apply
potential models in non-relativistic quantum mechanics to calculate
the mesons binding radius. Originally it was predicted that the
modification of the heavy quark pairs potential in the hot dense
matter created in heavy ion collisions would cause the pair to become
uncorrelated due to color charge screening. This modification of the
pair potential via a Debye mass term leads to charmonium
suppression~\cite{Matsui:1986dk} when compared to a binary collision
scaled p+p reference. Due to the different binding energies for the
different charmonium states one could gain access to the temperature
of the medium. At RHIC the suppression of the lowest energy charmonium
state, the $J/\psi$ meson, has been measured in Au+Au and Cu+Cu
collisions in $\sqrt{s}_{NN}=$200GeV collisions.

These measurements are then compared to the invariant yield measured
in a baseline p+p collision at the same center of mass energy. It is
assumed that any modification due to the medium will show up as a
deviation from the prediction of the binary collision scaled reference
data. This is of course true when no modifications due to normal
nuclear matter are present. However in the case of the $\jpsi$ we know
this is not the case from lower energy measurements made in p+A
collisions at SPS~\cite{Lourenco:2008sk} and
FNAL~\cite{Leitch:1999ea}.

Therefore, to interpret these data one must remove any effects that
occur in normal density cold nuclear matter (CNM). One such effect is
the modification of parton distribution functions in a nuclear
environment~\cite{deFlorian:2003qf,Eskola:2001gt}. Another is Cronin
enhancement that leads to a hardening of the transverse momentum
spectrum of collision products due to multiple scattering. To this end
the PHENIX experiment has also measured the nuclear modification
present in d+Au collisions in $\sqrt{s}=$200GeV collision. The d+Au
data is used to extrapolate within a Glauber~\cite{Miller:2007ri}
based data driven model to the Au+Au collision case to predict the
suppression that would result from CNM effects and search for an
anomalous suppression of $J/\psi$ mesons in the sQGP.

\section{$\jpsi$ Production}\label{}
\label{sec:production}
The production mechanism for charmonium has not been well understood
theoretically for nearly 20 years. The magnitude of the $p_{T}$
spectrum measured at the
Tevatron~\cite{Abulencia:2007us,Braaten:1994xb} was under-predicted by
an order of magnitude by the color singlet (CS)
model~\cite{Chang:1979nn}. This led to the proposal of the color octet
(CO) model~\cite{Bodwin:1994jh,Gong:2008ft} wherein the pre-charmonium
charm quark pair become correlated in a color charged state and must
color neutralize via soft gluon emission. While the CO model had some
success in describing the kinematic spectra it also predicted a large
transverse polarization at intermediate to large
$p_{T}$~\cite{Bodwin:1994jh,Gong:2008ft}. Another undesirable aspect
of the CO is its reliance on overlap matrices between the color
charged pre-hadron and final state J/$\psi$, which are nearly free
parameters in the model~\cite{Lansberg:2008gk}.

\begin{figure}[ht]
\centering
\includegraphics[width=0.45\textwidth]{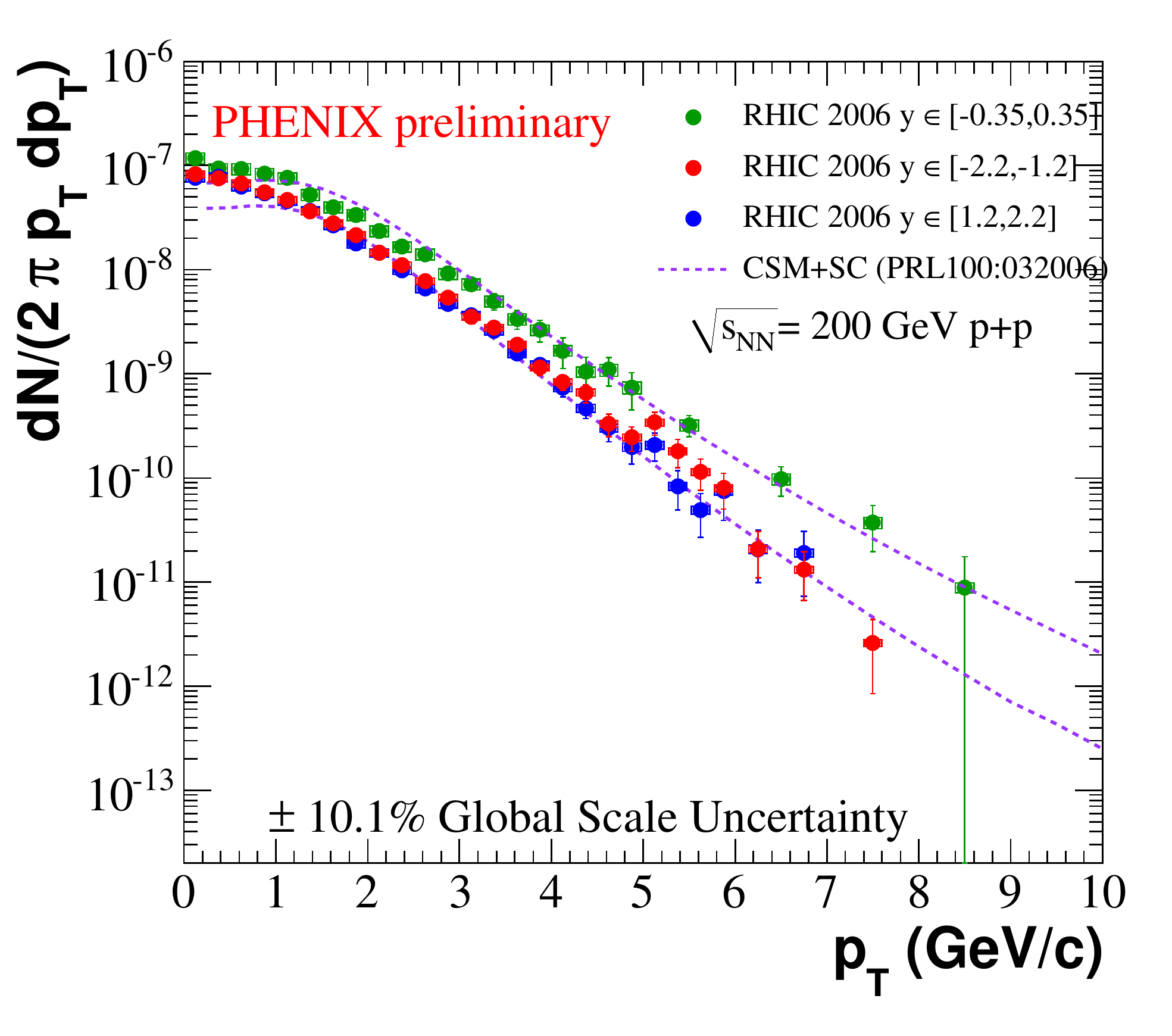}
\includegraphics[width=0.45\textwidth]{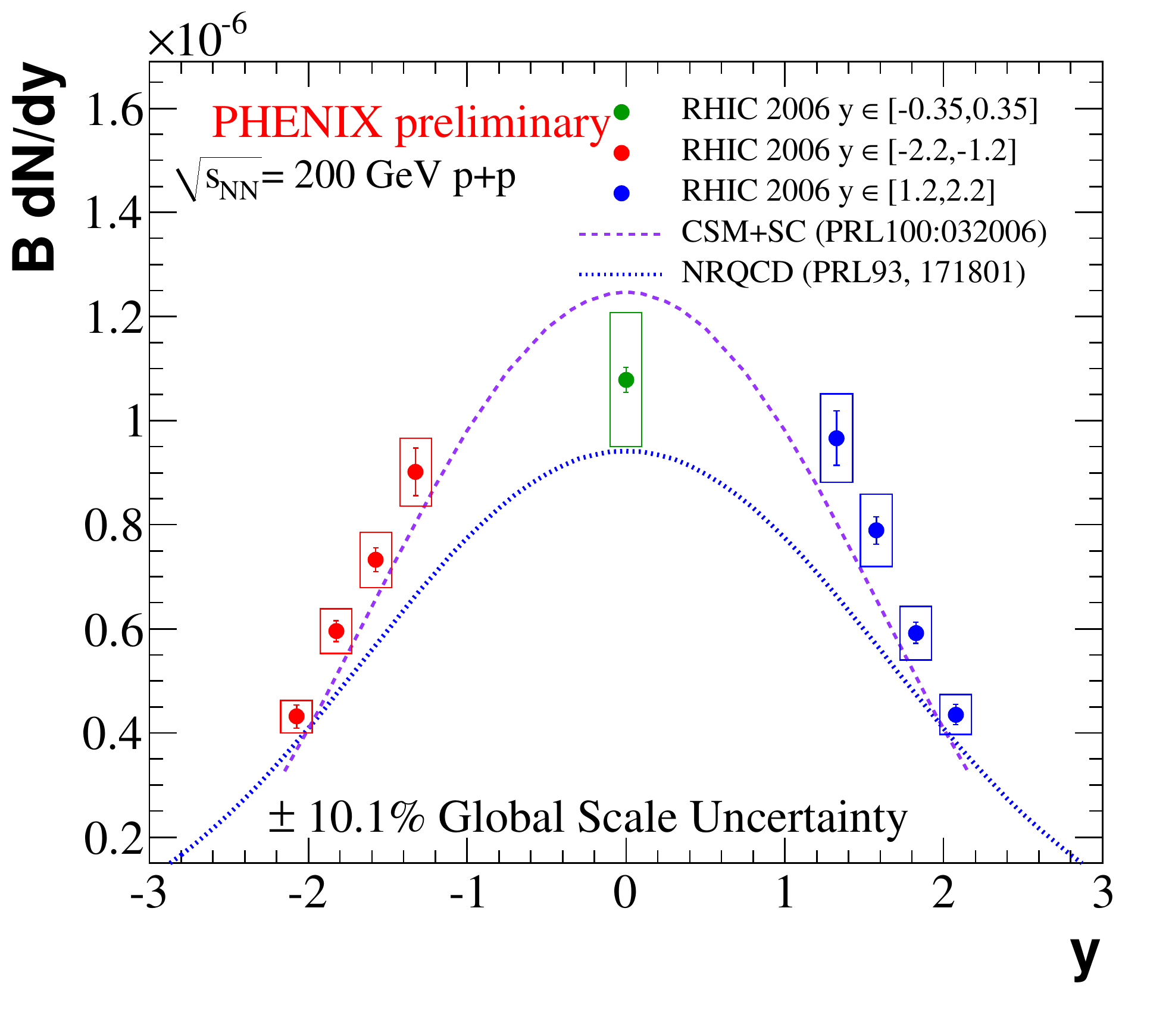} \\
\includegraphics[width=0.475\textwidth]{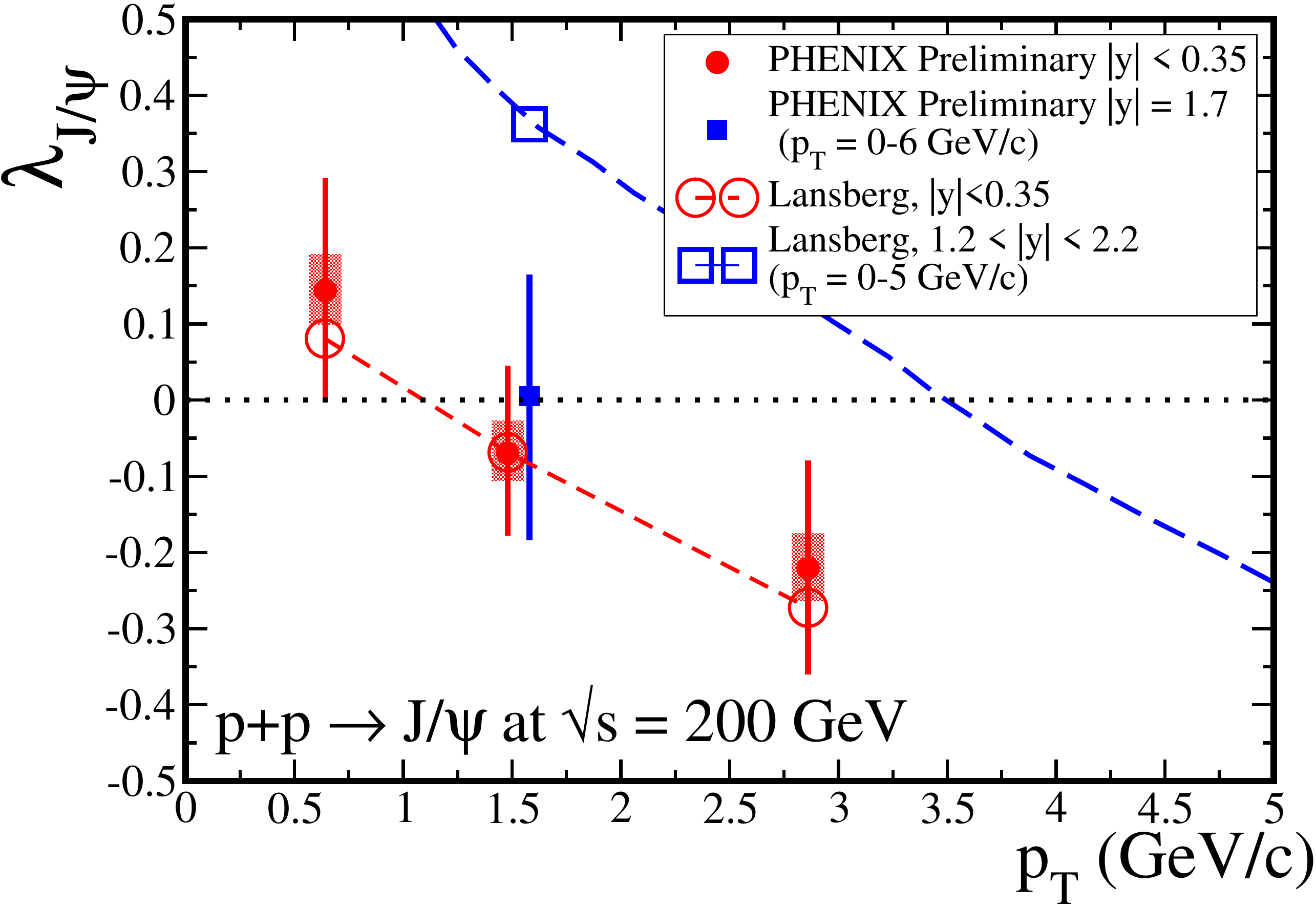}
\caption{\label{fig:invypp_polpp}(color online) Invariant yield of
  $\jpsi$ measured with the PHENIX spectrometer versus $p_T$ (top
  left) and y (top right). The theoretical curves are the four point
  modified CS model~~\cite{Lansberg:2008jn} and
  NRCD~\cite{Cooper:2004qe}. $\jpsi$ polarization versus $p_T$
  (bottom left) measured at mid and forward rapidity with the PHENIX
  spectrometer. The dashed lines are the predictions from the four
  point modified CS model~\cite{Lansberg:2008jn}.}
\end{figure}

Measuring the p+p spectra and polarization at RHIC is a tool for
distinguishing between charmonium production mechanisms.  PHENIX has
measured the invariant yield of J$/\psi$s, in p+p collisions at
$\sqrt{s}$ = 200 GeV, over a wide range in transverse momentum at
forward ($1.2 < \left|y\right| < 2.2$) and mid rapidity
($\left|y\right| < 0.35$) (Figure~\ref{fig:invypp_polpp} top) as well
as the J/$\psi$ polarization at mid rapidity
(Figure~\ref{fig:invypp_polpp} bottom). The quality of these data make
them natural metrics for testing new models of charmonium production
such as the four-point modified CS model proposed by Lansberg and
Haberzettl~\cite{Haberzettl:2007kj}.  The mid-rapidity data agrees
well with the polarization predictions. At forward rapidity there is a
two sigma difference between the measurement and prediction.

Another, quite recently measured, observable that may help establish
the production mechanism for the $\jpsi$ is azimuthal correlations
with hadrons. One could imagine that the spatial correlation between
the $\jpsi$ and the remainder of the hadrons forming the jet that
results from the hard initial collision may be very different
depending on the production mechanism. STAR has recently
released~\cite{Abelev:2009qa} azimuthal correlations between $\jpsi$s
and hadrons(Figure~\ref{fig:starcorr}) measured in $\sqrt{s_{NN}}$ =
200 GeV p+p collisions. The $\jpsi$ trigger particle is required to
have $p_T$ $>$ 5 GeV/c and the associated hadrons to have $p_T$ $>$
0.5 GeV/c (see Section~\ref{sec:feeddown}). This new observable may
prove very fruitful in distinguishing between different production
models in the future. One can imagine handing a MC generator
(i.e. PYTHIA) the QCD calculation for different models and extracting
the resulting correlation of prompt $\jpsi$ mesons and hadrons within
experimental cuts and then comparing these to data.

Understanding the production mechanism establishes the map between the
measured kinematics of the J/$\psi$ ($y$, $p_{T}$) and the kinematics
of the partons ($x_1$, $x_2$) at the interaction vertex. Different
production models provide these mappings and can result in different
conclusions about the magnitude of the cold nuclear matter
effects~\cite{Ferreiro:2008wc} discussed in Section~\ref{sec:cnm}. The
$J/\psi$ polarization and azimuthal correlation measurements provide
us with another lever arm to distinguish between these models.

\section{Feed Down}
\label{sec:feeddown}
One of the contributors to the J/$\psi$ spectrum is expected to come
from the feed down of higher charmonium states as well as $B$-meson
decays. The $\psi`$($\psi(2S)$), $\chi_c$ and $B$ mesons all have a
decay mode to J/$\psi$+X. Considering the predictions from the lattice
for the dis-association temperatures of these states at $\le T_C$
\cite{Mocsy:2007jz} (where $T_C \approx 170$ MeV~\cite{Adcox:2004mh})
and the most recent interpretations of the temperatures reached in
heavy ion collisions at RHIC $\approx 1.5T_C$ \cite{Adare:2008fqa} one
would expect the depletion of at least the $\chi_c$ and the $\psi`$
which would in turn lead to a J/$\psi$ suppression beyond that in cold
nuclear matter collisions. Of course this depletion must also be
accounted for when one attempts full accounting of J/$\psi$
suppression in heavy ion collisions.

To this end, the PHENIX collaboration has measured the feed down
fraction (R) of the $\psi'$ and set a 90\% confidence level upper
limit for the $\chi_c$ to J/$\psi$ in p+p collisions at 200 GeV via
the di-electron channel at central rapidity~\cite{Oda:2008kg}. The
results are in agreement with theoretical
predictions~\cite{Digal:2001ue} as well as the world
average~\cite{Faccioli:2008ir} (see table in~\cite{LindenLevysqm08}).

\begin{wrapfigure}{r}{0.5\textwidth}
\centering
\includegraphics[clip=true,viewport=0in 0in 7.75in 5in,width=0.5\textwidth]{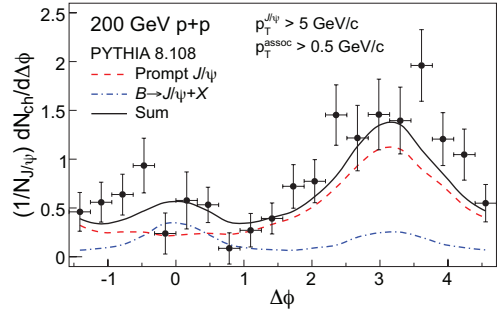}
\caption{\label{fig:starcorr} (color online) $\jpsi$-hadron azimuthal
  correlations measured with the STAR spectrometer.}
\end{wrapfigure}

As mentioned above the STAR experiment has also measured the azimuthal
correlation of $\jpsi$ and hadrons (Fig.~\ref{fig:starcorr}). Using
this correlation one can extract the feed-down fraction of $\jpsi$
arising from B meson decays if one knows the azimuthal shape of the
B-decay and prompt $\jpsi$ correlations. In order to extract this
quantity STAR parameterizes the shape as $C(\Delta{\phi}) =
x*C_{p}(\Delta{\phi}) + (1-x)*C_{B}(\Delta{\phi})$, where $C_{p}$ is
the correlation function from prompt $\jpsi$ mesons and $C_{B}$ comes
from the B decay. The prompt correlation function is then taken from a
PYTHIA simulation that has been tuned (i.e.both color octet and
singlet mechanisms mixed) to match real data transverse momentum
distributions. While this may be a rather drastic assumption, it is
necessary to extract the B feed down fraction in this manner. One
would naturally argue that an assumption of this nature would
inherently introduce a large systematic uncertainty. This is
especially true if one considers that the production mechanism (as
discussed above) is still not known and that PYTHIA simulations do not
reproduce either the cross section or polarizations measured in
data. However, this extraction yields a feed-down fraction of 13 $\pm$
5 \% \cite{Abelev:2009qa}.

\section{Cold Nuclear Matter Effects}
\label{sec:cnm}

In order to interpret suppression in heavy ion data one must remove
any effects that occur in normal density cold nuclear matter
(CNM). One such effect is the modification of parton distribution
functions in a nuclear
environment~\cite{deFlorian:2003qf,Eskola:2001gt}. Another is Cronin
enhancement that leads to a hardening of the transverse momentum
spectrum of collision products due to multiple scattering. The PHENIX
experiment has also measured the nuclear modification present in
deuteron+gold (d+Au) collisions in $\sqrt{s}$ = 200 GeV collision.

\begin{figure}[ht!]
  \centering
  \begin{minipage}[b]{0.49\textwidth}    
  \includegraphics[width=\textwidth]{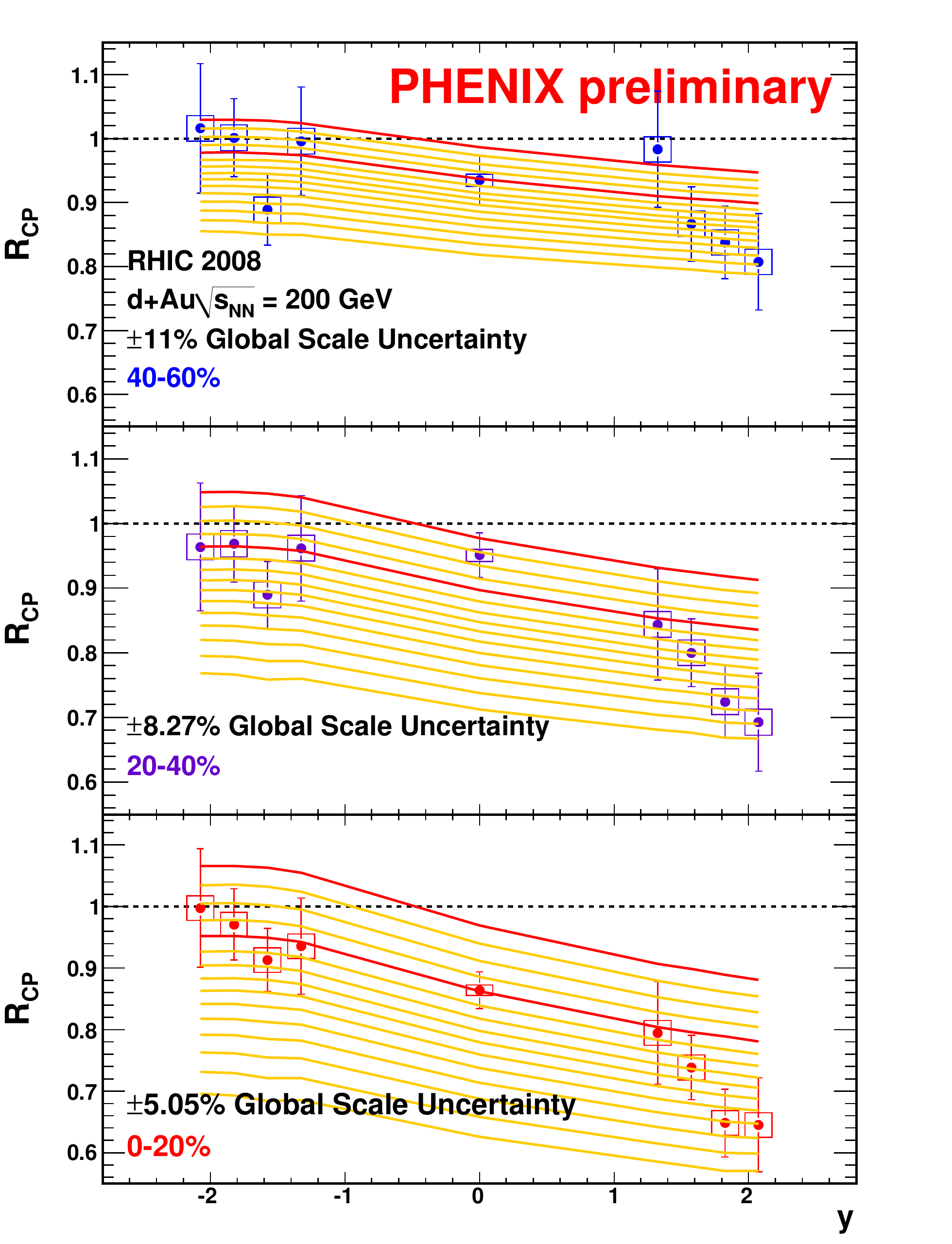}
  \end{minipage}
  \begin{minipage}[b]{0.49\textwidth}    
  \includegraphics[width=0.9\textwidth]{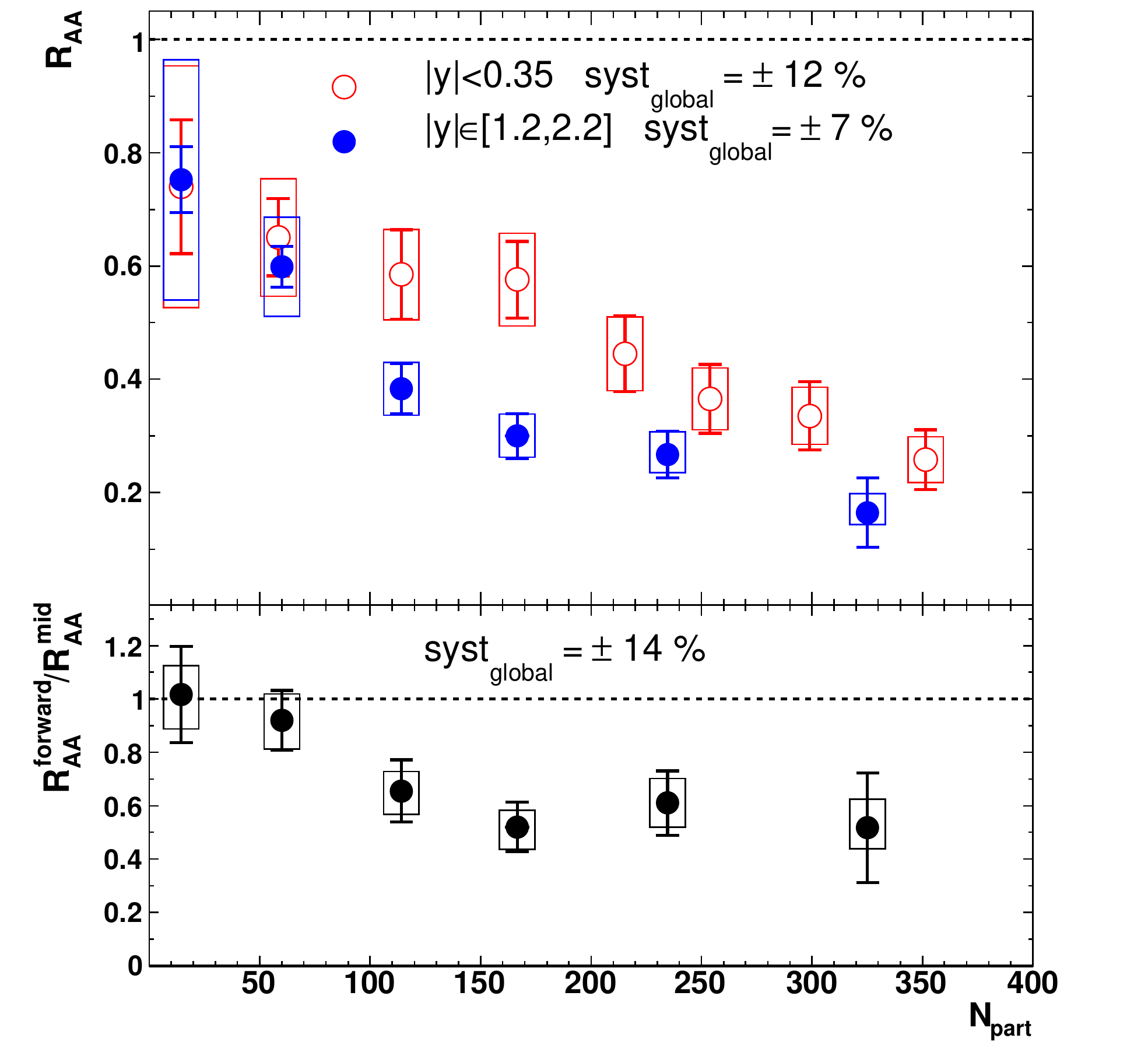}\\
  \includegraphics[width=0.95\textwidth]{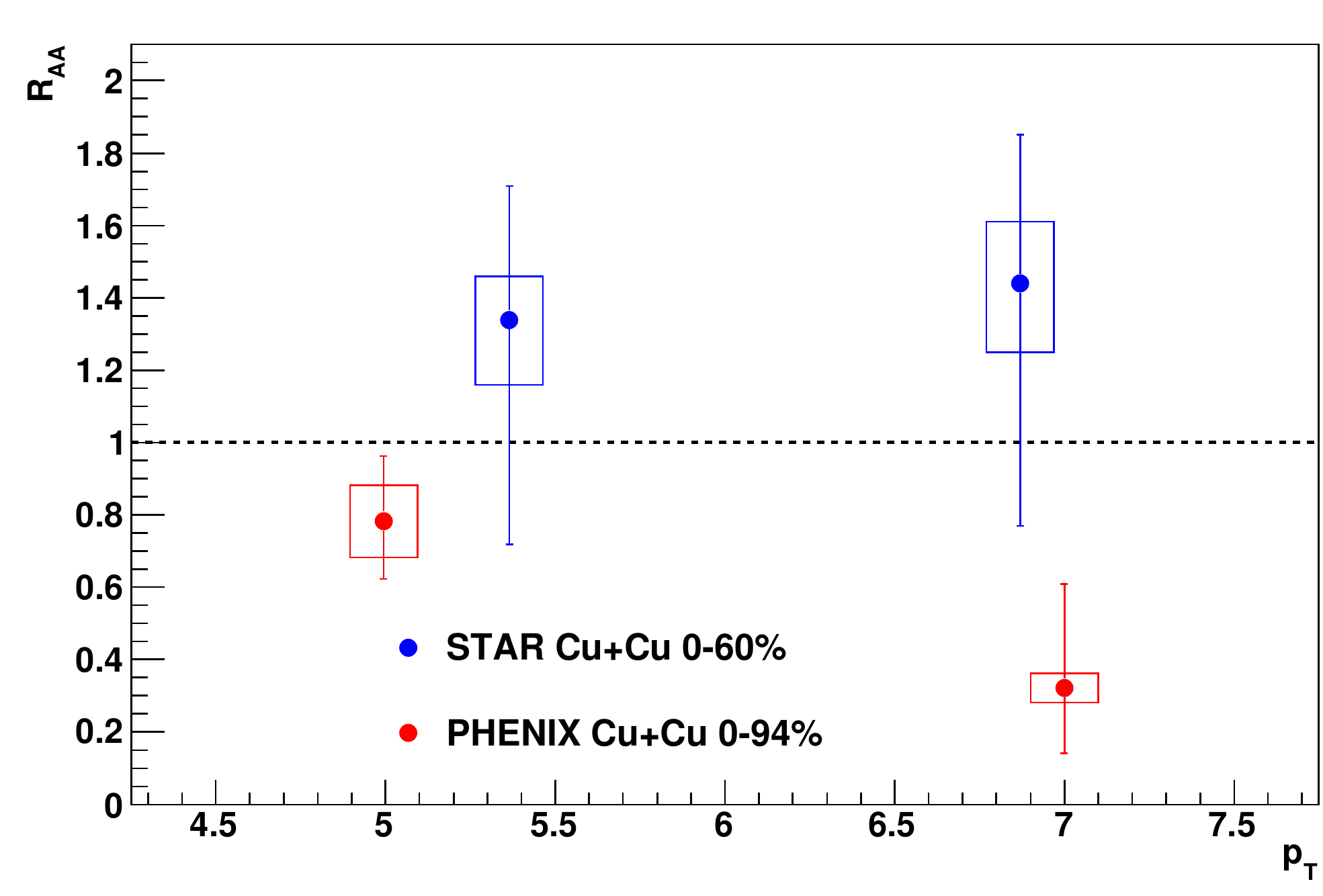}
  \end{minipage}
  \caption{\label{fig:run8rcp} (color online) Central to
    peripheral ratio ($R_{Cp}$) measured by the PHENIX collaboration
    using the RHIC 2008 d+Au data set (left). The theoretical curves
    represent the prediction for $R_{CP}$ using the
    EKS~\cite{Eskola:1999fp} nPDF parameterization and different
    constant-with-rapidity breakup cross sections. Published PHENIX
    Au+Au nuclear modification factor~\cite{Adare:2006ns} showing a
    larger suppression at forward rapidity than central (top
    right). Comparison of the PHENIX and STAR results for nuclear
    modification in Cu+Cu at $\sqrt{s}=200$ GeV (bottom right).}
\end{figure}

During the 2008 RHIC deuteron-gold (d+Au) run PHENIX recorded a factor
of 30 greater integrated luminosity than the previous run in 2003. The
PHENIX collaboration has not yet released nuclear suppression factors
for d+Au collisions compared to the Run-3 p+p baseline. However, the
data were analyzed to measure the central to peripheral ratio $R_{CP}$
in these collisions (Eq.~\ref{eq:rcp}). In order to calculate the
central to peripheral ratio $R_{CP}$ the invariant yield of $J/\psi$
mesons from d+Au central and peripheral collisions must be
measured. Figure~\ref{fig:run8rcp} shows the nuclear modification as a
function of rapidity ($y$) in a given centrality bin ($i$) which
defines the average number of binary collisions ($N^{coll}_{i}$).

These data have sparked much interest in the community (for details
see the 2009 ECT and INT quarkonia workshops) as it is clear that a
rapidity independent breakup cross section combined with nuclear
shadowing cannot match the shape of the data. %
This has lead to the conclusion that there may be some physics missing
in the models. PHENIX is currently working to understand a
normalization effect, between different RHIC runs, for the nuclear
suppression factor ($R_{dAu}$) and these results are expected to be
made available later this year.

\begin{equation}
\centering R_{CP}(y;i) = \frac{dN^{d+Au}_{i}/dy}{N^{coll}_{i}
  dN^{d+Au}_{periph}/dy }
\label{eq:rcp}
\end{equation}


\section{What's Hot in Hot nuclear matter}
\label{hotnuc}
In Figure~\ref{fig:run8rcp} (top right) we show the nuclear suppression factor
for Au+Au collisions at $\sqrt{s}$ = 200 GeV, as a function of
centrality, measured at forward and mid rapidity by the PHENIX
collaboration at RHIC~\cite{Adare:2006ns}. One striking feature of
this comparison is the similarity between the suppression patterns at
mid rapidity between the RHIC ($\left|y\right| < 0.35$) and SPS ($0 <
y < 1$) data~\cite{Scomparin:2007rt}, despite the difference in center
of mass energies of the two measurements ($\sqrt{s}$ = 200 GeV and
$\sqrt{s} \sim$  20 GeV respectively). This engenders the question ``Why
are the results so similar?'' for measurements probing different
rapidity regions, shadowing regimes and energy densities. It is clear
from the PHENIX data that the suppression at forward rapidity is
greater than at mid rapidity. This is a challenge to local density
based suppression models and is one piece of evidence that supports
the idea of regeneration discussed below, where the $\jpsi$ yield is
enhanced due to close proximity in phase space of uncorrelated pairs.


We also show (Figure~\ref{fig:run8rcp} (bottom right)) the high $p_T$
measurements of $R_{AA}$ made by the PHENIX and STAR experiments in
Cu-Cu at $\sqrt{s_{NN}}$ = 200 GeV. One should note that the PHENIX
measurement is for minimum bias collisions while the STAR measurement
is for centrality 0-60\%. Given this small difference one can still
compare the measurements. It is clear that the STAR data favors a
reduction of the suppression at high $p_T$ à(with large systematic and
statistical uncertainty), while the PHENIX data favors a nearly
constant suppression from mid to high $p_T$.

\section{A glimpse of the future.}\label{}
\label{future}
Regeneration (also called
coalescence)~\cite{BraunMunzinger:2000px,Thews:2000rj} is a process
whereby un-correlated $c\bar{c}$ pairs coalesce to form charmonium
states, an effect that could be enhanced due to the close phase space
proximity of partons in a heavy ion collision. This process would
increase the yield of J/$\psi$ in heavy ion collisions and must be
accounted for when interpreting the overall magnitude of suppression.

\begin{figure}
\centering
\includegraphics[width=0.50\textwidth]{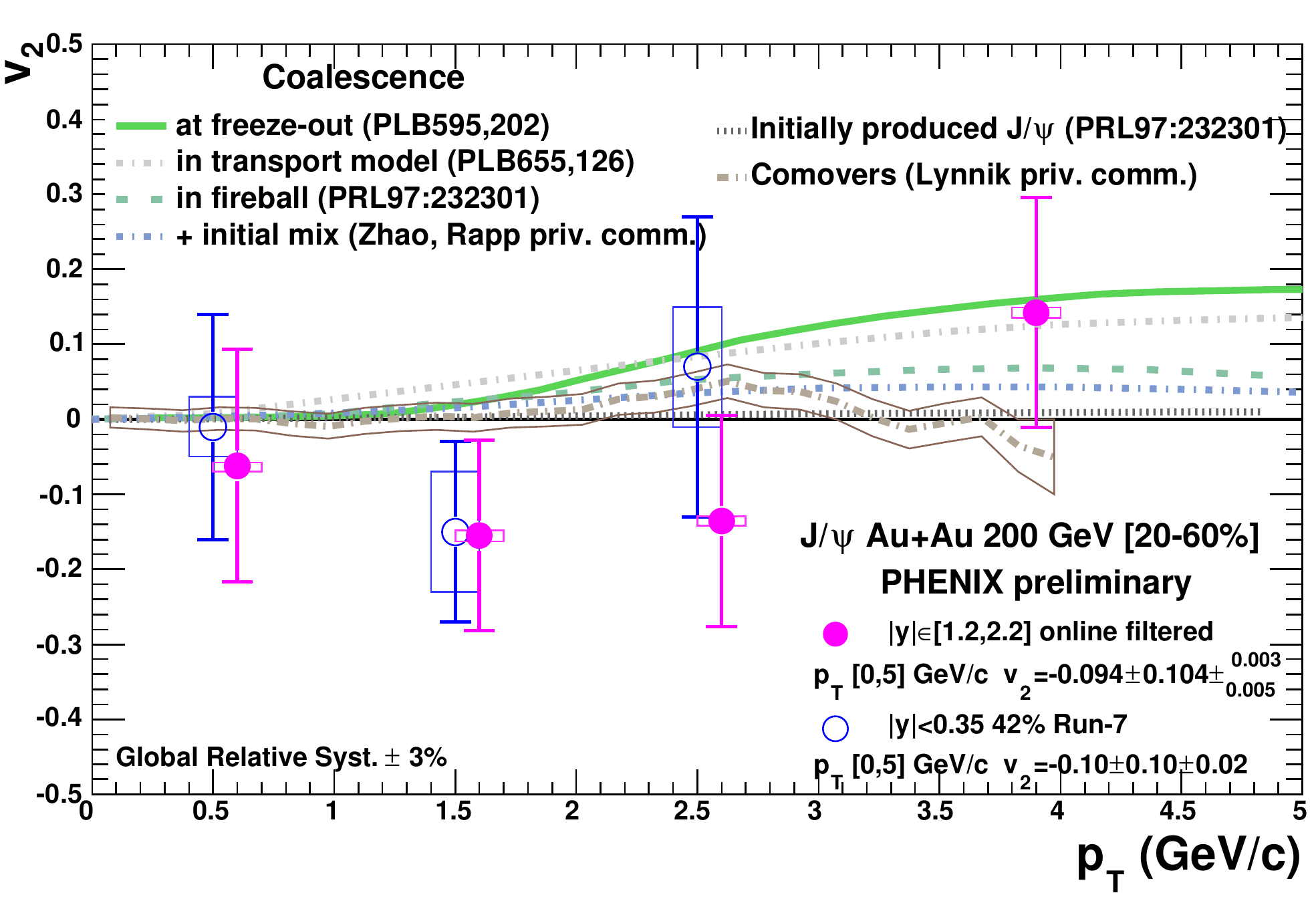}
\caption{\label{run7v2} (color online) Elliptic flow ($v_2$) versus
  $p_T$ measured with the PHENIX detector at mid and forward rapidity
  compared to various model
  predictions~\cite{Greco:2003vf,Zhao:2007hh,Ravagli:2007xx,Yan:2006ve,Linnyk:2008uf}.}
\end{figure}

It has been suggested that one metric for determining the magnitude
and/or existence of this effect would be to measure the elliptic flow
($v_2$) of the J/$\psi$ in heavy ion collisions. The elliptic flow of
the recombined J/$\psi$ would be related to the flow of uncorrelated
charm~\cite{Molnar:2003ff} in the medium and could be very different
from that of J$\psi$ formed early on in the fireball's evolution. This
experimental signature can be compared to the $v_2$ from open charm
and models that contain the regeneration mechanism. The PHENIX
experiment has made the first measurement of J$/\psi$ elliptic flow at
forward and mid rapidity~\cite{Silvestre:2008nk} for RHIC
energies. The result does not have the statistical precision to
distinguish between models and the comparison to the open charm
elliptic flow~\cite{Adare:2006nq} is not very enlightening
(Figure~\ref{run7v2}). A long, high-luminosity Au+Au run at RHIC will
be necessary for this metric to be improved.

\begin{figure}[h!]
\centering
\includegraphics[width=0.45\textwidth]{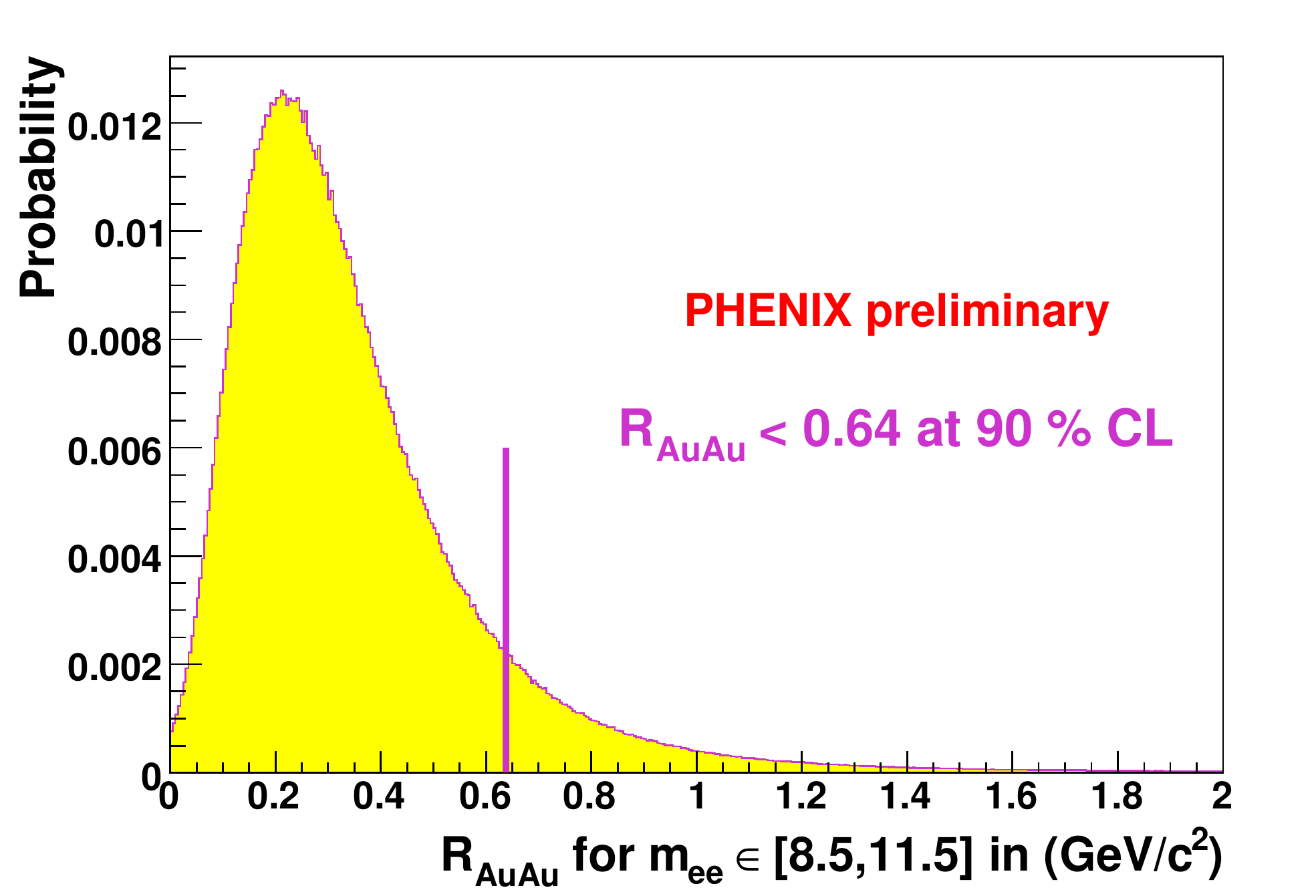}
\includegraphics[width=0.35\textwidth]{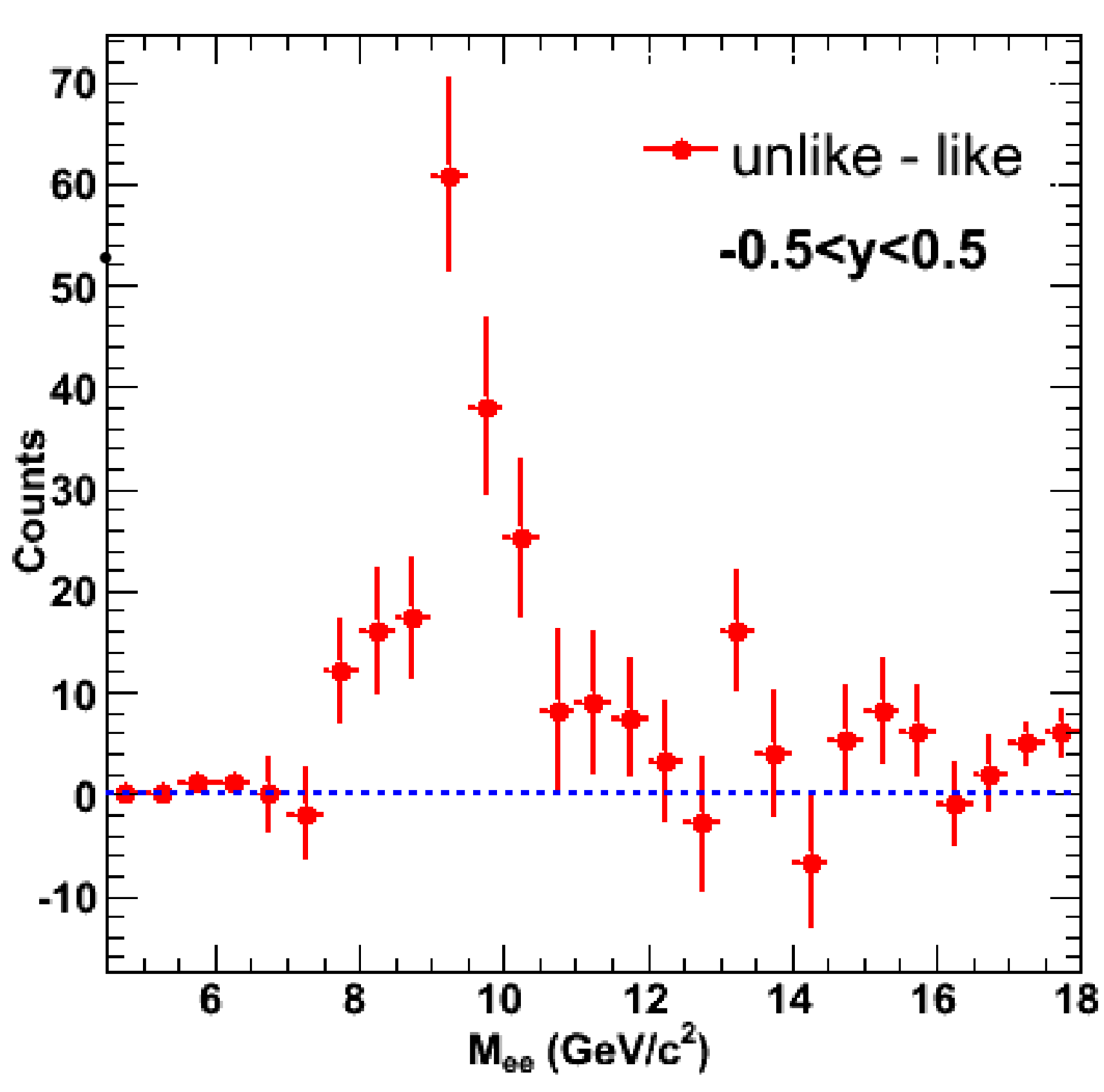}
\caption{\label{fig:upsilon} (color online) Nuclear modification
  probability distribution from PHENIX for $\Upsilon$ suppression in
  Au+Au collisions (left). Di-electron invariant mass spectrum from
  STAR using the Run-8 d+Au data that shows a prominent $\Upsilon$
  peak (right).}
\end{figure}

In addition STAR has measured a clear $\Upsilon$ peak in the d+Au data
taken in 2008 Figure~\ref{fig:upsilon} (right). This coupled with the
previous p+p measurement~\cite{Djawotho:2007} can be used to calculate
$R_{dAu}$ for the $\Upsilon$. The value is 0.98 $\pm$ 0.32 (stat.)
$\pm$ 0.28 (syst,). This measurement suffers from very large
uncertainties which engenders the question of whether or not the RHIC
program for higher mass charmonium states would benefit from a long
d+Au run in the future. The current measurement would allow for
suppression in cold nuclear matter of up to ~30\% within one standard
deviation of the statistical error alone. A long d+Au run will be
required if one wants to precisely characterize any cold nuclear
matter effects that would be present for the $\Upsilon$ state.

PHENIX has also released a new measurement of $\Upsilon$ suppression
in Au+Au collisions. Due to a lack of statistics the value is
presented as an upper limit on the amount of suppression seen in a HI
collision. An upper limit of 0.64, at a 90\% confidence level, is
calculated by coupling the Poisson probability distributions for the
numerator (Au+Au) and denominator (p+p) to arrive at a probability
distribution for $R_{AuAu}$ Figure.~\ref{fig:upsilon}
(left). Interpreting this result is difficult due to the lack of
information regarding cold nuclear matter effects for the
$\Upsilon$. However, note that if the $\Upsilon$(2s) and
$\Upsilon$(3s) states are melted in the medium a suppression of ~30\%
would be expected in HI collisions.



\bibliographystyle{elsarticle-num}
\bibliography{phenix}

   

\end{document}